\documentclass[12pt]{article}

\usepackage{cite}
\usepackage{epsf}

\def\beq{\begin{equation}}
\def\eeq#1{\label{#1}\end{equation}}
\def\eeqn{\end{equation}}
\def\beqa{\begin{eqnarray}}
\def\eeqa#1{\label{#1}\end{eqnarray}}
\def\eeqan{\end{eqnarray}}

\def\leqn#1{(\ref{#1})}

\def\stacksymbols #1#2#3#4{\def\theguybelow{#2}
    \def\vp{\lower#3pt}
    \def\sp{\baselineskip0pt\lineskip#4pt}
    \mathrel{\mathpalette\intermediary#1}}

\def\intermediary#1#2{\vp\vbox{\sp
     \everycr={}\tabskip0pt
     \halign{$\mathsurround0pt#1\hfil##\hfil$\crcr#2\crcr
              \theguybelow\crcr}}}

\def\gapproxeq{\stacksymbols{>}{\sim}{2.5}{.2}}
\def\lapproxeq{\stacksymbols{<}{\sim}{2.5}{.2}}

\begin{document}

\begin{titlepage}
\begin{flushright}
Saclay T02/038\\
LBNL-50102 \\
{\tt hep-ph/0204142} \\
\end{flushright}

\vskip.5cm
\begin{center}
{\huge{\bf Fine Structure Constant Variation}}
\vskip.4cm
{\huge{\bf  from a Late Phase Transition}}
\vskip.2cm
\end{center}
\vskip0.2cm

\begin{center}
{\sc Z. Chacko}$^{a,b}$,
{\sc C. Grojean}$^{c}$
and {\sc M. Perelstein}$^{a}$

\end{center}
\vskip 10pt

\begin{center} $^{a}$ {\it Theory Group, Lawrence Berkeley National 
Laboratory, Berkeley, CA 94720, USA } \\ 
\vspace*{0.1cm}
$^{b}$ {\it Department of Physics, University of California, Berkeley, CA 94720, USA } \\
\vspace*{0.1cm}  
$^{c}$ {\it Service de Physique Th{\'e}orique, CEA--Saclay,
F--91191 Gif--sur--Yvette, France} \\
\vspace*{0.1cm}
{\tt zchacko@thsrv.lbl.gov, grojean@spht.saclay.cea.fr, meperelstein@lbl.gov}
\end{center}

\vglue 0.3truecm

\begin{abstract} 
\vskip 3pt 
\noindent 
Recent experimental data indicates that the fine structure constant $\alpha$
may be varying on cosmological time scales. We consider the possibility that 
such a variation could be induced by a second order phase transition which 
occurs at late times (z $\sim$ 1 - 3) and involves a change in the vacuum
expectation value (vev) of a scalar with milli-eV mass.  Such light scalars 
are natural in supersymmetric theories with low SUSY breaking scale. If
the vev of this scalar contributes to masses of electrically charged fields, 
the low-energy value of $\alpha$ changes during the phase transition. 
The observational predictions of this scenario include isotope-dependent
deviations from Newtonian gravity at
sub-millimeter distances, and (if the phase transition is a sharp event
on cosmological time scales) the presence of a well-defined step-like
feature in the $\alpha(z)$ plot. The relation between the fractional
changes in $\alpha$ and the QCD confinement scale is highly model dependent,
and even in grand unified theories the change in $\alpha$ does
not need to be accompanied by a large shift in nucleon masses.
\end{abstract}

\end{titlepage}
\newpage

\setcounter{footnote}{0}
A recent analysis~\cite{Webb3} of quasar absorption spectra at redshift
$z\sim 0.5$--$3.5$ indicates that the fine structure constant $\alpha$ is 
changing with time:
\beq
\frac{\alpha_\mathrm{now}-\alpha_\mathrm{past}}{\alpha_\mathrm{now}}
\equiv \frac{\delta \alpha}{\alpha} = (.71 \pm .18) \, 10^{-5}.
\eeq{change}
While theorists have considered the possibility that the fundamental 
constants are time-dependent for a long time (starting with 
Dirac~\cite{Dirac}), it is not clear how the result \leqn{change} fits into 
the 
current field-theoretic picture of elementary particle physics. In the
Standard Model (SM), all the coupling constants run with energy. Since
the temperature of the Universe changes with
time, this implies that the effective values of the couplings are also
changing. This effect, however, cannot be used to explain the data, since 
$\alpha$ does {\it not} run at energies below 
the electron mass $m_e \approx 500$ keV, corresponding to redshifts of 
order $10^7$. Thus, it appears that the time
variation of $\alpha$ reported in~\cite{Webb3} cannot be explained without
appealing to physics beyond the Standard Model.

A priori, string theory should be well suited to accommodate this data. 
Indeed, it is well known that in string theory any coupling constant
is promoted to a vacuum expectation value (vev) of a scalar field such as 
the dilaton or some other modulus. If this scalar field 
is extremely light, $m\sim10^{-33}$~eV, its expectation value could be still 
evolving in the recent past (or even today.) However, it is not clear how
such a low scalar mass can be stabilized against radiative corrections.
Moreover, the superlight scalar will mediate a long-range, isotope-dependent 
force. Such forces are subject to severe constraints from 
accurate tests of the equivalence principle (for a recent review, 
see~\cite{Damour}) and other fifth force experiments~\cite{Bek, OP, Dvali}. 
Finally, it has been pointed out~\cite{Dine} that the energy density 
associated with the rolling scalar field is large enough to overclose the 
Universe if the parameters of the model are chosen to accommodate the result
\leqn{change}. (This problem seems to be tightly connected with the usual 
cosmological constant problem, and may become a non-issue once that
problem is resolved~\cite{LSS}.)  

In this paper we propose an alternative explanation of the time variation of 
the fine structure constant that is potentially experimentally distinguishable
from the slowly rolling dilaton scenario. 
Our framework is based on the observation that the experiments~\cite{Webb3} 
measure the value of the fine structure constant at the very low energy scale 
(about 10 eV) set by atomic physics. This infrared value of the constant,  
which we will denote by $\alpha\equiv\alpha(0)$, is related to its 
ultraviolet value $\alpha(\Lambda)$ by the renormalization group 
equation:
\beq
\frac{1}{\alpha} = 
\frac{1}{\alpha (\Lambda)} 
+ \frac{1}{2\pi} \sum_{i=0}^{N} b_{i+1} \ln \frac{m_{i+1}}{m_i},
\eeq{rg}
where $m_0<m_1<m_2<\ldots<m_N$ are the masses of all electrically
charged particles, commonly referred to as ``thresholds'' ($m_0=m_e$, 
$m_1=m_\mu$, etc.) and $b_i$ is the one--loop beta function 
between the scale $m_{i-1}$ and $m_i$. For notational convenience, we have 
identified $m_{N+1}$ with the cutoff of the theory $\Lambda.$
With our conventions, $b<0$ for an asymptotically free theory. It is
clear from \leqn{rg} that the low-energy value of $\alpha$ can change even if
its fundamental, short-distance value $\alpha (\Lambda)$ remains fixed,
provided that some of the thresholds move. It is this possibility that 
we would like to investigate in this paper. 

A change in $\alpha$ induced by a small variation of thresholds is easily
obtained from \leqn{rg}:
\beq  
\frac{\delta \alpha}{\alpha} = 
-\frac{\alpha}{2 \pi} \sum_{i=0}^{N} (b_i-b_{i+1}) \frac{\delta m_i}{m_i}.
\eeq{deltas}
In QED, $b_i-b_{i+1}=-q_{i}^2 n_i<0$, where
$q_i$ is the electric charge of the particle that decouples at $m_i$, and
$n_i$ is its degeneracy. To induce the change of $\alpha$ at $10^{-5}$ 
level, we need
\beq
\sum_i \frac{\delta m_i}{m_i} \,\sim\, 10^{-2}.
\eeq{deltam}
Notice that the masses have to increase ($m_\mathrm{now}-m_\mathrm{past}>0$)
in order to reproduce
the correct sign in the evolution observed by Ref.~\cite{Webb3}.
The requirement (\ref{deltam}) is not in obvious contradiction with other 
observations. While the bounds on the variation of electron and 
proton masses at $z \sim 2 -3$ are rather 
tight~\cite{Potekhin}, there are no constraints on the 
masses of other charged particles ({\it e.g.} muon or $\tau$) at these 
redshifts.

What physical mechanism could lead to a change in the mass of a charged
particle? In many theories masses of charged fermions are determined by
vacuum expectation values of electrically neutral scalar fields. (For  
example, lepton and quark masses in the Standard Model are proportional to the
Higgs vev.) In the expanding Universe, scalar vevs can be time-dependent.
In particular, we will consider models in which in the early Universe the
vev of a certain scalar $S$ vanishes due to thermal effects. As the Universe 
cools down, a phase transition occurs, during which the scalar acquires a 
vev. As a result, the masses of electrically charged states coupled to $S$
will change, leading to a change in the low-energy value of $\alpha$ according
to \leqn{deltas}.

Let us determine the features of the zero-temperature scalar field potential 
$V(S)$ which are necessary to explain the data~\cite{Webb3}. We assume 
that in the early Universe $S = 0$. It is easy to see that $V^\prime(0)$ has 
to vanish; however, $S = 0$ could be either a minimum or a (local) 
maximum of $V(S)$. In the latter case, the vacuum can still be stable,
provided that a thermal mass term is generated by interactions of the 
$S$ field with the surrounding plasma~\cite{Weinberg}: 
\beq
V_{\rm therm} \sim T^2 S^2. 
\eeq{thermal}
For the
moment, we will simply assume that such a mass term is generated; we will
later comment on the conditions under which this is the case. As the 
temperature drops to its critical value, characterized by 
\beq
T_c^2 \sim -V^{\prime\prime} (0),  
\eeq{Tc}
the vacuum at $S=0$ becomes unstable, and the system undergoes a phase 
transition. In this transition, the field changes its vev until
it reaches the nearest minimum of $V(S)$, which we will denote by $S_1$. 
We require that the phase transition occurs at $z \sim 1 - 3$, when the 
temperature of the Universe $T$ is of order $10^{-3}$ eV. Eq.~\leqn{Tc}
then implies that the mass parameter of the $S$ field in the 
zero-temperature lagrangian is of the same order, $10^{-3}$ eV. In 
a generic non-supersymmetric theory, such a low mass scale is unstable
against radiative corrections, and extreme fine tuning would be required to
explain it. In supersymmetric theories, however, this scale can be 
radiatively stable, provided that the supersymmetry breaking scale 
$\sqrt{F}$
is at or below about 10 TeV and the breaking is only communicated to $S$ via 
Planck suppressed contact operators\footnote{The fact that the energy scale 
corresponding to the current critical density of the Universe could arise 
naturally in supersymmetric theories was emphasized in Ref.~\cite{Hitoshi}.}.
Such a low value of $\sqrt{F}$ 
is phenomenologically viable if the breaking is communicated to the 
visible sector fields (that is, the Standard Model fields and their 
superpartners) by gauge mediation~\cite{Ann}. Thus, we have 
identified a large class of models in which a phase transition could occur  
at low redshifts, without any fine-tuning.

The phase transition will not result in a change of the fine structure
constant unless $S$ is coupled to at least one electrically charged 
field. On the other hand, radiative stability of the $S$ mass parameter
would be destroyed by any direct renormalizable coupling of $S$ to the 
Standard Model fields or their superpartners. To avoid this problem, we 
introduce an additional
vectorlike chiral superfield $Q$, which is charged under $U(1)_{\rm em}$
(and possibly other SM gauge groups.) This field is coupled to $S$ via
\beq
{\cal L} = \int d^2\theta \, (M\,+\,yS) Q \bar{Q},
\eeq{couple} 
where $M$ is a mass for the $Q$ and $\bar{Q}$ fields, and $y \sim 1$ 
is a Yukawa coupling.
It is easy to see that three-loop diagrams involving $Q$ and $\bar{Q}$ 
renormalize the $S$ mass parameter by an amount
\beq
(\delta m)^2 \sim \left({y^2 \over 16\pi^2}\right)\,\left({e^2 \over 16\pi^2}
\right)^2 \, {F^2 \over M^2}.
\eeq{threeloop}
These corrections do not destroy radiative stability provided that 
$M \gapproxeq 10^{16}$ GeV. Note that
Eqs. \leqn{deltam} and \leqn{couple} imply that $S_1/M \sim 10^{-2}$ is
necessary to accommodate the data, so
the vev of the $S$ field after the transition is required to be very large, 
$S_1 \gapproxeq 10^{14}$ GeV. The huge hierarchy between the vev of the
field and its mass is radiatively stable because of supersymmetry
and implies that $S$ is a modulus.

Let us return to the question of whether a thermal mass term of the form
\leqn{thermal} is actually generated at high temperatures.
Since the couplings of $S$ to the visible sector fields are non-renormalizable,
they cannot generate such a term. Moreover, 
since $S$ is a modulus, its self-interactions are extremely weak: the
quartic coupling can be estimated as $\lambda \sim m^2/S_1^2 \sim 10^{-52}$. 
The thermal mass term generated by these interactions is proportional to 
$\lambda$ and is therefore negligible. To generate a thermal mass  
of the right order of magnitude, $S$ needs to have substantial couplings
to some other light 
fields $Y$. These couplings will {\it not} 
destabilize the $10^{-3}$ eV mass scale, provided that the fields $Y$,
like $S$ itself, 
are only sensitive to supersymmetry violating effects through Planck 
suppressed operators. (Together with $S$, these fields form a 
``hidden sector'' of the theory.) Although the fields of the
hidden sector decouple from the visible sector fields very early on, they
can maintain thermal equilibrium among themselves, at a temperature which is
identical (up to order one factors related to the multiplicity of states at
decoupling) to the visible sector temperature. The interactions of $S$
with this ``hidden'' thermal plasma are responsible for generating the mass
term~\leqn{thermal}.

The natural value of the energy
density difference before and after the phase transition is given by 
\beq
\Delta V \equiv V(0)-V(S_1) \sim |V^{\prime\prime} (0)| \,S_1^2
\eeq{deltaV}
and given the above constraint
we obtain $\Delta V > 10^{40}$ eV$^4$.  Naively, such a huge energy density 
seems to lead to two severe problems, which would make our scenario 
incompatible with
standard cosmology. First, one could argue that the cosmological constant 
cannot be tuned away both before and after the phase transition, and rapid
inflationary expansion should occur in at least one of these two periods.   
Second, during the phase transition the field $S$ is expected to undergo 
coherent oscillations, slowly decaying due to Hubble expansion. With our
parameters, the energy density in these oscillations is large enough to 
overclose the Universe. Requiring that our scenario {\it without any 
additional ingredients} be consistent with
standard cosmology leads to a
bound on the variation of $\alpha$ similar to the one found in 
Ref.~\cite{Dine}: $\delta\alpha/\alpha\lapproxeq 10^{-31}$, well below 
the values reported in~\cite{Webb3}. 

A simple way to restore the consistency of our scenario with standard 
cosmology is to simply fine-tune the shape of the potential. Such fine tuning
can be used to get rid of the large contribution to the vacuum energy before
the phase transition, Eq.~\leqn{deltaV}. Once this is done, the energy of
coherent oscillations is automatically low enough to avoid overclosure. 
The amount of fine tuning involoved is similar to what is required in the
conventional models with a rolling dilaton field~\cite{Dine}.

It is tempting, however, to speculate that the difficulties of our scenario 
could be avoided
without any fine tuning once the cosmological constant problem is resolved.
Indeed, from the effective
field theory point of view, it is reasonable to expect that the mechanism 
that solves the cosmological constant problem will guarantee the absence of 
inflation {\it regardless} of the (constant) vacuum energy 
density\footnote{This point of view was advocated, for example, in 
Ref.~\cite{selftuning}.}. During a phase transition, such a 
mechanism would adjust itself to cancel the new value of the energy density.
Any mechanism which has this feature would resolve the first of the two
problems confronting our scenario. Furthermore, the energy in the coherent 
oscillations of the modulus field is related to the difference in the
vacuum energies before and after the transition. It is the large value of
this difference that leads to the second problem of our scenario. If the
large difference in vacuum energies is cancelled by the adjustment mechanism,
conservation of energy implies that the amount of energy in the coherent
oscillations will be sufficiently small to avoid overclosure.
Of course, we emphasize that no explicit example of an adjustment mechanism 
for the cosmological constant problem is currently known, and therefore
the above discussion is necessarily rather speculative. If such a 
mechanism is proposed in the future, the question of whether our scenario
is viabile without fine tuning should be reexamined within a more concrete 
framework.   

\if
The prediction of large-amplitude coherent oscillations of the 
modulus field relies on the assumption that the dynamics of the field during 
the transition is not influenced by the adjustment mechanism. However, 
if adjustment occurs on sufficiently short time scales, the buildup of
coherent oscillations could be prevented. Whether this actually happens 
depends on the details of the adjustment mechanism. Thus, we believe 
that at the present time our scenario cannot be rigorously ruled out by 
cosmological considerations.  
Of course, since no adjustment mechanism for the cosmological constant 
problem is currently known, the scenario remains incomplete. If such a 
mechanism is proposed in the future, the viability of the scenario should be 
reexamined within a more concrete framework.   
\fi

In the context of a grand unified theory a change in $\alpha$ would seem to
require a variation of the QCD confinement scale. This, in turn, would result 
in a shift of the hadron masses. In fact, if the observed variation of 
$\alpha$ is attributed to the change of the short-distance unified coupling 
$\alpha_{\rm GUT}$, the fractional change in the proton mass can
be predicted~\cite{CF, LSS} (see also~\cite{DF} for an estimate
in the change of the deuteron binding energy) 
and turns out to be about 40 times larger than the
fractional change in $\alpha$. This result is in mild contradiction with the 
bound obtained from measuring the value of $\mu=m_e/m_p$~\cite{Potekhin} in
the same range of redshifts, $z \sim 2-3.$ In our framework, however, it is 
straightforward to evade this prediction. This possibility is perfectly 
compatible with grand unification. For example, consider a supersymmetric
 SO(10) grand
unified theory in which doublet-triplet splitting is achieved using an adjoint
field $\Sigma$ with block diagonal vacuum expectation value 
$(0,0,0,\tau_2,\tau_2)$ where $\tau$s are the Pauli matrices~\cite{DP,CM}. 
This pattern of vevs breaks SO(10) down to SU(4) $\times$ SU$(2)_L\times$ 
U(1$)_{I3R}$, where the U(1) is the diagonal generator of SU(2$)_R$. The
superpotential of the theory necessarily contains a Planck suppressed operator
of the form $S \Sigma Q 
\bar{Q}/ M_{Planck}$, where $Q$ and $\bar{Q}$ are in the fundamental of 
SO(10), and $S$ is a singlet. Because of the pattern of vevs of 
the field $\Sigma$, the singlet $S$ acquires Yukawa couplings of the form
\leqn{couple} to the doublets in $Q$ and $\bar{Q}$ but not the triplets.
When $S$ goes through a phase transition, the low energy values of the
SU(2$)_L$ and U(1$)_Y$ gauge couplings are affected, but the SU(3) gauge
coupling (and therefore the QCD confinement scale) remains unaffected
at one loop order, and the resulting change in the proton mass will 
therefore be suppressed. However, there will be changes in both the 
electron mass and the proton mass 
arising from radiative corrections to these quantities involving  
electromagnetic and weak interactions. 
Moreover, since supersymmetry breaking is mediated to the visible sector 
fields through gauge interactions, the Higgs mass and therefore its vacuum 
expectation
value will be altered by the change in the values of the  SU(2) and U(1) gauge 
couplings. Both these effects will in turn give rise to a change in $\mu$
of order $\delta\mu/\mu \sim \delta\alpha/\alpha$. Although 
this is below the current experimental bound, it may be possible to detect 
this effect in future experiments.

The observational predictions of our model depend on the time $T$ it takes
to complete the phase transition. Since the dynamics of the transition is
necessarily strongly influenced by the unknown adjustment mechanism for 
cosmological constant, it is not possible to estimate $T$ reliably. Let
us consider two cases. In the first case, $T$ is smaller than the age of 
the Universe at the time when the transition began, so that the transition
is a sharp event on cosmological time scales. 
Then, the low-energy value of the fine structure
constant should be time-independent for $z \lapproxeq 1$. This prediction 
is consistent with the null results from the laboratory search~\cite{Clocks} 
for the time variation of $\alpha$, as well as the geological
bound~\cite{Oklo} from the Oklo natural nuclear reactor. Moreover, in 
this case we would expect improved measurements of $\alpha(z)$ to show
a sharp feature (a ``step'') in the $z\sim 1-3$ region. The other
possibility is that the time $T$ is so large that the transition is
still not completed today.  
In this case, it is much harder to distinguish
our scenario from the more conventional picture of a slowly rolling
dilaton field. 

Apart from the quasar absorption spectra measurements, the only 
currently available bound on the 
variation of $\alpha$ on cosmological time scales comes from 
nucleosynthesis~\cite{BBN}; our model easily satisfies this
constraint ($|\Delta \alpha|/\alpha<10^{-2}-10^{-4}$).
The expected precision~\cite{CMB} in the cosmic microwave background
measurements ($|\Delta \alpha|/\alpha<10^{-2}-10^{-3}$)
will not be enough to rule out our scenario. 

An observational consequence of the rolling dilaton mechanism of 
changing $\alpha$ is the existence of a new, isotope-dependent 
long-range force~\cite{Bek, OP, Dvali}. In our scenario, the modulus $S$ 
(in the true 
vacuum) has a mass of about $10^{-3}$ eV and mediates a Yukawa force  
with a range of order 0.1 mm. Since there are no renormalizable
couplings between $S$ and any of the Standard Model fields, this force
is of (approximately) gravitational strength. Currently, the strongest bounds
on such forces come from tabletop precision tests of the Newton's 
law at submillimeter distances~\cite{Adelberger}. These experiments 
require that the range of the extra force be less than about 0.2 - 0.3 mm.
Clearly, this bound can be satisfied in our model without significant fine 
tuning. If our scenario is indeed realized in nature, improved precision
gravitational tests should observe isotope-dependent deviations from the 
Newton's law in the near future. 

Above, we have assumed that $S=0$ is a local maximum of the zero-temperature
potential $V(S)$, and the phase transition is second order.
It is also possible that $S=0$ is a local minimum of the potential. If there
is another minimum with lower energy, at $S=S_1$, a first order phase
transition could occur. During this phase transition, bubbles of the
true vacuum nucleate and start growing rapidly. Eventually, the bubbles
coalesce and the transition is complete. A classic calculation (neglecting
the effects of the expansion of the Universe) of the lifetime of a false
vacuum~\cite{Coleman} yields
\beq
\tau \sim \Lambda \exp \left(
\frac{27\pi^2}{8} \,\frac{S_0^4}{(\Delta V)^3}
\right)
\eeq{tunnel} 
where $\Lambda$ is the natural scale of the potential $V(S)$,
$\Delta V = V(0) - V(S_1)$ is the energy splitting between the true and
false vacua, and
\beq
S_0 = \int_{0}^{S_1} dS \sqrt{V(S)} \,\sim\, S_1 \sqrt{\Delta V}.
\eeq{action}
The phenomenological constraints on this scenario are somewhat different
from the case of a second order transition discussed above. According to
\leqn{tunnel}, the lifetime of the false vacuum can be much larger than
$\Lambda$ without major fine tuning.
Thus, it is possible for a late phase transition to occur even if all the
scales in $V(S)$ are of order TeV or higher. This in turn allows direct
couplings of $S$ to the (electrically charged) visible sector fields. Note,
however, that the absence of rapid inflationary expansion before (or after)
the phase transition would naively require $\Delta V <(10^{-3}$ eV$)^4$.
This bound would not only necessitate fine tuning of the energy splitting
itself, but also lead to unacceptable large values of $\tau$ (much larger
than the age of the Universe today) unless the entire shape of $V(S)$
is finely tuned. On the other hand, just like in the case of second order
phase transition discussed earlier, this bound could be nullified once the
physics responsible for solving the cosmological constant is taken into
account.

To summarize, we have suggested that the apparent time variation of the
low energy fine structure constant reported in~\cite{Webb3} is a result
of a late phase transition involving changing the vacuum expectation
value of a light modulus. We have shown that such a late phase transition
can occur in a large class of supersymmetric models without any fine
tuning. This is in marked contrast with a more conventional rolling dilaton
mechanism of changing $\alpha$, which requires superlight (Hubble-scale)
moduli whose mass is generally not radiatively stable even in the absence
of gravity. Just like in the
rolling dilaton picture, producing a change in $\alpha$ large enough to
explain the data~\cite{Webb3} in our scenario is challenging, and
naively seems incompatible with cosmological observations. In both pictures,
this problem could be avoided by fine tuning the potential of the scalar 
field. We have speculated that our scenario {\it could} be compatible with 
standard cosmology without fine tuning, provided that the cosmological 
constant problem is resolved by an adjustment mechanism. Clearly, this
question would have to be reexamined if an explicit adjustment mechanism
is proposed in the future. While we are
not able to reliably describe the dynamics of the phase transition, we 
emphasize that the transition could be a sharp event on cosmological 
time scales. In this case, improved measurements of the redshift dependence 
of $\alpha$ will discover a well-defined step-like feature, distinguishing 
this scenario from the rolling dilaton picture. Another prediction 
is the isotope-dependent deviation from Newtonian gravity at
sub-millimeter scales. Finally, we have found that in our framework the 
relation between the change in 
$\alpha$ and the corresponding change in the QCD confinement scale is highly 
model dependent. In particular, even in supersymmetric grand unified 
theories it is straightforward to construct models which are consistent with 
current bounds on the variation of the ratio of the electron mass to the 
proton mass at $z \sim 2-3$.  

\vskip0.7cm
\noindent
{\bf Acknowledgments}\\
\noindent

We thank C.~Cs\'aki, E.~Dudas, N.~Kaloper, M.~A.~Luty, H.~Murayama,
A.~E.~Nelson and J.P.~Uzan for fruitful discussions.
M.P. wishes to acknowledge the hospitality of Service de Physique Th\'eorique,
CEA--Saclay, where this work was initiated.
Z.C. and M.P. are
supported by the Director, Office of Science, Office of High
Energy and Nuclear Physics, of the U. S. Department of Energy
under Contract DE-AC03-76SF00098.


\end{document}